\documentclass[prb,showpacs,preprintnumbers,amsmath,amssymb,superscriptaddress]{revtex4}
\usepackage{epsf}
\usepackage{graphicx}
\usepackage{bm}% bold math
\usepackage{color}

\begin{document}

\title{Tunneling magnetic effect in heterostructures with paramagnetic impurities}

\author{I.~V.~Rozhansky}
\email{rozhansky@gmail.com} \affiliation{A.F.~Ioffe Physical
Technical Institute, Russian Academy of Sciences, 194021
St.Petersburg, Russia} \affiliation{Lappeenranta University of
Technology, P.O. Box 20, FI-53851, Lappeenranta, Finland}
\author{N.~S.~Averkiev}
\affiliation{A.F.~Ioffe Physical Technical Institute, Russian
Academy of Sciences, 194021 St.Petersburg, Russia}
\author{E.~Lahderanta}
\affiliation{Lappeenranta University of Technology, P.O. Box 20,
FI-53851, Lappeenranta, Finland}


\date{\today}

\begin{abstract}
An effect of paramagnetic impurity located in a vicinity of a
quantum well (QW) on spin polarization of the carriers in the QW is
analyzed theoretically. Within approach of Bardeen's tunneling
Hamiltonian the problem is formulated in terms of Anderson-Fano
model of configuration interaction between a localized hole state at
Mn and continuum of heavy hole states in the InGaAs-based QW. The
hybridization between the localized state and the QW leads to
resonant enhancement of interband radiative recombination. The
splitting of the configuration resonances induced by splitting of
the localized state in magnetic field results in circular
polarization of light emitted from the QW. The developed theory is
capable of explaining known experimental results and allows for
calculation of the photoluminescence spectra and dependence of
integral polarization on temperature and other parameters.
\end{abstract}

\maketitle

%\index{Rozhansky I. V.}
%\index{Averkiev N. S.}
\section{Introduction}
Various phenomena based on interference of a discrete
quantum-mechanical state and a continuum of states have been
intensively studied since the famous paper by U. Fano
\cite{PhysRev.124.1866} rated among the most relevant works of 20th
century \cite{RevModPhys.82.2257}. He suggested a theoretical
approach often regarded as Fano-Anderson model or Fano configuration
interaction which was successfully applied to explain the
characteristic asymmetric resonances observed in atomic spectroscopy
experiments. It further appeared that numerous examples of Fano
resonances existed in atomic and nuclear physics, condensed matter
physics and optics \cite{RevModPhys.82.2257}. The co-existence of a
localized discrete level and continuum of states within the same
energy range is also quite common in low-dimensional semiconductor
structures
\cite{RevModPhys.82.2257,PhysRevB.65.155302,okulov:220,springerlink:10.1134/S1063782608080034}.
The structures containing a quantum well (QW) and a ferromagnetic or
paramagnetic layer located in the vicinity of the QW nowadays are
gaining a special interest as they are believed to combine the high
mobility of the carriers in the QW and magnetic properties provided
by the magnetic layer. In particular, an exchange interaction with
ferromagnetic layer leads to spin polarization of
holes\cite{springerlink:10.1134/1.1641485}. For GaAs--based
structures with a $\delta$--layer of Mn the holes probably play an
important role in promoting ferromagnetic state of the Mn
layer\cite{0953-8984-20-14-145207,10.1063/1.3267314}. The system
considered in the present work consists of GaAs--based
heterostructure with a In$_{x}$Ga$_{1-x}$As QW ($x=0.1-0.2$) and
$\delta$--layer of paramagnetic acceptor (Mn) located at a distance
of a few nanometers from the QW. A number of recent experiments show
that the Mn $\delta$--layer gives rise to circular polarization of
the photoluminescence from the QW in magnetic field
\cite{springerlink:10.1134/S1063783410110144,springerlink:10.1134/S0021364009220056}.
It was found that the intensities of the radiation at the wavelength
corresponding to interband direct transitions in the QW differ for
opposite circular polarizations in magnetic field applied normally
to the QW plane. If Mn is replaced by a non-magnetic acceptor
(carbon) no polarization is registered. Thus the observed
polarization is not due to g-factor of the free carriers in the QW
which in this way is proved to be small. On the contrary, the holes
localized at Mn do have
 $g\approx3$ \cite{PhysRevLett.59.240} and possibly can penetrate into the QW
 by means of quantum-mechanical tunneling which is expected to be of
 resonant--type if the energy of the localized state coincides with that of free 2D hole in the QW.
We use the Fano configuration interaction approach to study
tunneling hybridization between a localized hole state at
paramagnetic impurity with the 2D continuum of states in the QW. We
show how this hybridization should reveal itself in the
photoluminescence at the QW wavelength. The theory allows for
calculation of circular polarization degree in magnetic field. We
have to mention here that there are other mechanisms that might
contribute to the observed polarization like non-resonant tunneling
of electrons from QW to Mn and recombination with the holes
localized at Mn. While it still remains unclear which of the
mechanisms has larger contribution to the polarization in
experiments on photolumenescence
\cite{springerlink:10.1134/S1063783410110144,springerlink:10.1134/S0021364009220056},
in our paper we restrict our consideration by configuration
interaction for the holes quite naturally expected in the p-type
system under study.

\section{Hole states at acceptor and in quantum well}
\label{secStates} We assume that the eigenstates of the holes in the
QW form a continuum of states starting from the energy of size
quantization (taken for zero of energy scale). These 2D states can
be characterized by two quantum numbers, say $k_x$ and $k_y$ are the
projections of 2D wavevector on the axis $x$ and $y$ in the plane of
the QW. Considering interaction with a point-like impurity it is
more convenient use cylindrical basis. In this case each state in
the QW is characterized by the magnitude of the wavevector $k$ and
index $l$ denoting the cylindrical harmonic. The corresponding
wavefunction is given by:
\begin{equation}
\label{eqwavefuncylindr} \varphi_{kl}\left(\rho,z\right)={\eta
\left( z \right)\frac{{{\pi ^{1/4}}}}{{\sqrt 2 }}\frac{{\sqrt k
}}{{{S^{1/4}}}}{J_l}\left( {k\rho } \right){e^{il\theta }}},
\end{equation}
where $J_l(k\rho)$ is Bessel function of order $l$, $\rho$ and
$\theta$ are the polar coordinates in the plane of the QW, $S$ is
the area of the QW plane, $\eta \left( z \right)$ is the envelope
function of size quantization in $z$-direction. The wavefunction is
normalized to unity. Firstly, let us treat this continuum as a set
of discrete states characterized by energies $\varepsilon_{k}$ and
wavefunctions $\varphi_{kl}$. Considering the free 2D carriers
implies the magnetic field applied to the QW to be nonquantizing.
The validity of this assumption is discussed in section
\ref{secDiscus}. Below we consider In$_{x}$Ga$_{1-x}$As QW having
only one level of size quantization for the heavy holes, we neglect
the light holes being split off due to size quantization. Thus the
basis of Bloch amplitudes to be used is formed of the states with
certain projection of total angular momentum $J=3/2$ on $z$ axis
which is perpendicular to the QW plane:
%($J=3/2$)
\begin{equation} \label{eqBasis} \left(
{{e_{3/2}},\,\,{e_{1/2}}\,,\,{e_{ - 1/2}}\,,\,{e_{ - 3/2}}}
\right).\end{equation} The wavefunctions $\varphi_{kl,j}$ in this
basis have the form:
\begin{equation}
\label{eqwavefunlqw} {\varphi _{kl, - \frac{3}{2}}} = \left(
{\begin{array}{*{20}{c}}
   0  \\
   0  \\
   0  \\
   {{\varphi _{kl}}\left( {\bf{\rho }} \right)}  \\
\end{array}} \right),\,\,\,\,{\varphi _{kl, + \frac{3}{2}}} = \left( {\begin{array}{*{20}{c}}
   {{\varphi _{kl}}\left( {\bf{\rho }} \right)}  \\
   0  \\
   0  \\
   0  \\
\end{array}} \right).
\end{equation} The kinetic energy of a state in the QW is related to
the wavevector value as \[{\varepsilon} = \frac{{{\hbar
^2}{k^2}}}{{2m'_{hh}}},\] where $m_{hh}'$ is the in-plane heavy hole
mass in the QW.

In order to determine wavefunction $\psi$ of a hole localized at an
acceptor one should consider the kinetic part of the Luttinger
Hamiltonian and attractive potential of the acceptor $U(r)$. The
spherically symmetrical potential preserves the symmetry $\Gamma_8$,
thus the ground state is 4-fold degenerate and can be classified by
projection of angular momentum. The eigenfunctions of Luttinger
Hamiltonian with spherically symmetric attractive potential can be
explicitly found in the model of zero radius potential
\cite{Averkiev}. In the basis of Bloch amplitudes they are expressed
as follows:
\begin{eqnarray}
\label{eqWavefuncions} \psi_{+\frac{3}{2}} = \left(
{\begin{array}{c}
%3/2
   {{R_0}{Y_{00}} + \frac{1}{\sqrt 5}{R_2 Y_{20}}}  \\
   { - \frac{2}{{\sqrt {10} }}{R_2}{Y_{21}}}  \\
   {\frac{2}{{\sqrt {10} }}{R_2}{Y_{22}}}  \\
   0  \\
\end{array}} \right), \;
%1/2
\psi_{+\frac{1}{2}} = \left( {\begin{array}{c}
   {\frac{2}{{\sqrt {10} }}{R_2}{Y_{2, - 1}}}  \\
   {{R_0}{Y_{00}} - \frac{1}{{\sqrt 5 }}{R_2}{Y_{20}}}  \\
   0  \\
   {\frac{2}{{\sqrt {10} }}{R_2}{Y_{22}}}  \\
\end{array}} \right), \nonumber \\
%-1/2
\psi_{-\frac{1}{2}} = \left( {\begin{array}{c}
   {\frac{2}{{\sqrt {10} }}{R_2}{Y_{2, - 2}}}  \\
   0  \\
   {{R_0}{Y_{00}} - \frac{1}{{\sqrt 5 }}{R_2}{Y_{20}}}  \\
   {\frac{2}{{\sqrt {10} }}{R_2}{Y_{21}}}  \\
   \end{array}} \right),\;
%-3/2
\psi_{-\frac{3}{2}} = \left( {\begin{array}{c}
   0  \\
   {\frac{1}{{\sqrt 5 }}{R_2}{Y_{2, - 2}}}  \\
   { - \frac{2}{{\sqrt {10} }}{R_2}{Y_{2, - 1}}}  \\
   {{R_0}{Y_{00}} + \frac{1}{{\sqrt 5 }}{R_2}{Y_{20}}}  \\
\end{array}} \right). \end{eqnarray}
Here \begin{align}
 {R_0} &= {C_0}\left( {\frac{\beta }{r}{e^{ - qr\sqrt \beta  }} + \frac{{{e^{ - qr}}}}{r}} \right), \nonumber \\
 {R_2} &= {C_0}\left( {\frac{\beta }{r}{e^{ - qr\sqrt \beta  }}\left( {1 + \frac{3}{{qr\sqrt \beta  }} + \frac{3}{{{q^2}{r^2}\beta }}} \right) - \frac{{{e^{ - qr}}}}{r}\left( {1 + \frac{3}{{qr}} + \frac{3}{{{q^2}{r^2}}}} \right)} \right), \nonumber \\
 {C_0} &= \sqrt {\frac{q}{{{\beta ^{3/2}} + 1}}},  \nonumber \\
 q &= \sqrt {\frac{{2m_{hh}{E_0}}}{{{\hbar ^2}}}},  \nonumber \\
 \beta  &= \frac{{{m_{lh}}}}{{{m_{hh}}}},
 \end{align}
$E_0$ is the binding energy of a hole at the acceptor, $Y_{lm}$ are
the spherical harmonics. $m_{lh}, m_{hh}$ - respectively are the
bulk masses of light hole and heavy hole in GaAs.
 Note, that the radial part of all nonzero components of the
wavefunctions (\ref{eqWavefuncions}) have two characteristic decay
lengths, the largest of the two being always determined by the light
hole mass  $m_{lh}$.

\section{Tunneling between acceptor and quantum well for complex band structure}
\label{secT} For the studied system the potential barrier separating
the Mn $\delta$--layer from the QW is weakly transparent for
tunneling so the localized hole state mix up with the QW states.
Rigorous calculation of eigenfunctions of the resulting state is
rather hard to perform as it requires solving stationary Schrodinger
equation in the complicated 3D potential and with account for the
complex valence band structure. In order to circumvent the explicit
solving of Schrodinger equation for tunneling problems the so-called
tunneling or transfer Hamiltonian formalism is commonly used as
originally proposed by Bardeen \cite{PhysRevLett.6.57}. Within this
approach the total tunneling Hamiltonian is expressed as
$H=H_{Mn}+H_{QW}+H_T$, where $H_{Mn}$ is partial Hamiltonian having
the localized hole state at Mn as its eigen state. $H_{QW}$ in the
same way corresponds to the QW itself, its eigenfunctions
$\varphi_k$ form non-degenerate continuum of states characterized by
quantum number $k$. The term $H_T$ accounts for the tunneling. In
the secondary quantization representation the total Hamiltonian can
be written as follows:
\begin{equation}
\label{eqTunHamil} H = {\varepsilon_0}{a^ + }a + \sum\limits_k
{{\varepsilon _k}c_k^ + {c_k} + } \sum\limits_k {{T_k}c_k^ + a +
T_k^*{a^ + }{c_k}},
\end{equation} where $a^+, a$ --
the creation and annihilation operators for the localized state
characterized by its energy $\varepsilon_0$, and $c^+_k, c_k$ -- the
creation and annihilation operators for the continuum state having
energy $\varepsilon _k$. For the convenience both energies
$\varepsilon_0$ and $\varepsilon _k$ here and below are measured
from the level of size quantization of holes in the QW so that
$\varepsilon _k$ is simply their kinetic energy.
 The expression (\ref{eqTunHamil}) is rather general, in fact
it can be regarded as introduction of the coupling between two
systems into the Hamiltonian in the most simple phenomenological
way. From this viewpoint the coupling parameter $T_k$ is still to be
determined through exact solving of the eigenvalue problem for the
whole system. Bardeen's approach suggests a simple recipe for
calculation of the tunneling parameter $T_k$ (also referred as
tunneling matrix element) for the case of weak tunneling through
potential barrier:
\begin{equation}
\label{eqTunel} {T_k} = \left\langle {{\varphi_k}K{\psi}}
\right\rangle - \left\langle {{\psi}K{\varphi_k}} \right\rangle^* ,
\end{equation}
where $K$ is the kinetic energy operator, commonly equal to
\[K =  - \frac{{{\hbar ^2}}}{{2m}}\Delta .\]
For this problem the 3D integration (\ref{eqTunel})
straightforwardly reduces to integration along any surface inside
the barrier. In order to apply the same approach
%$known to be reliable in numerous cases
to the problem of the holes tunneling in GaAs it has to be
generalized for the case of complex band structure. It would be
tempting to do it by treating $K$ in (\ref{eqTunel}) as the kinetic
part of the effective Hamiltonian and $\psi$, $\varphi_k$ as
envelopes in the basis of Bloch amplitudes. For the valence band of
In$_{x}$Ga$_{1-x}$As -- based heterostructure the holes ground state
with total angular momentum $J=3/2$ is described by Luttinger
Hamiltonian ($\hbar k_x$,$\hbar k_y$,$\hbar k_z$ are, as usual, the
momentum operators along the appropriate axis):
\begin{equation} \label{eqLuttinger} K = \left(
{\begin{array}{*{20}{c}}
   F & H & I & 0  \\
   {{H^*}} & G & 0 & I  \\
   {{I^*}} & 0 & G & { - H}  \\
   0 & {{I^*}} & { - {H^*}} & F  \\
\end{array}} \right),
\end{equation}

\begin{align}
 F &=  - A{k^2} - \frac{B}{2}\left( {{k^2} - 3k_z^2} \right), \nonumber \\
 G &=  - A{k^2} + \frac{B}{2}\left( {{k^2} - 3k_z^2} \right), \nonumber \\
 H &= D{k_z}\left( {{k_x} - i{k_y}} \right) \nonumber, \\
 I &= \frac{{\sqrt 3 }}{2}B\left( {k_x^2 - k_y^2} \right) -
 iD{k_x}{k_y}.
 \end{align}
The functions $\psi$, $\varphi_{k}$ in (\ref{eqTunel}) become
4-component vector functions in the basis (\ref{eqBasis}).
Analogously to the simple band case integration over the whole space
is reduced to the integration over the surface $\Omega_S$ inside the
barrier, at that, only $z$--projection of the kinetic energy
operator is required. For the eigenfunctions in the QW with zero
light hole components the expression for tunneling parameter
simplifies into:
\begin{equation}
\label{eqTLuttinger} {T_k} = \left( {B - A} \right)\int_{\Omega_S}
{dS} \left( {{\varphi _{k}}^*\frac{d}{{dz}}{\psi} - {\psi
}\frac{d}{{dz}}{\varphi _{k}}^*} \right).
\end{equation}
Regrettably, this straightforward generalization of (\ref{eqTunel})
fails to be correct. Indeed, the decay length of wavefunctions
(\ref{eqWavefuncions}) is determined by light hole mass while the
decay length of the of the QW states (\ref{eqwavefunlqw}) is
governed by the heavy hole mass. Due to this circumstance the result
of the surface integration (\ref{eqTLuttinger}) becomes crucially
dependent on the particular position of the integration surface
inside the barrier. However, it can be shown that in the case of two
masses the exponential dependence of the tunneling parameter $T_k$
on the barrier thickness is determined by the smallest mass. The
value of $T_k$ is also governed by a formfactor corresponding to the
overlap of the wavefunctions along the QW plane. We consider the
kinetic energy of a hole in the QW being small compared to the QW
depth, i.e. $k<<q$. In this case the overlap of the localized
wavefunction occur only with the zeroth harmonic of
(\ref{eqwavefuncylindr}). The only nonzero tunneling matrix elements
are therefore the following:
\[\begin{array}{l}
 {T_k} = {T_1}_k + {T_2}_k \\
 {T_1}_k = \left( {A - B} \right)\int_{\Omega_S} {dS}  \left( {{R_0}{Y_{00}}{J_0}\left( {k\rho } \right)\frac{d}{{dz}}\eta \left( z \right)} \right), \\
 {T_2}_k = \left( {A - B} \right)\int_{\Omega_S} {dS} \left( {\frac{1}{{\sqrt 5 }}{R_2}{Y_{20}}{J_0}\left( {k\rho } \right)\frac{d}{{dz}}\eta \left( z \right)} \right), \\
 \end{array}\]
The calculation of these integrals presents no difficulty for any
particular surface inside the barrier, however further detalization
of these values is beyond the accuracy of the approach. The only
statement valid is that the tunneling parameter $T_k$ has an
exponential dependence on the barrier thickness with the light hole
mass entering the exponent index:
\begin{align}
\label{eqT}
% & T  =  T_1  + T_2  \nonumber \\
 & T_k  =  \left( {A - B} \right)\zeta  \left( {k/q}
\right)\frac{{\pi ^{3/4} q\sqrt k }}{{S^{1/4} }}\exp \left( { - \chi
 \left( {k/q} \right)qd\sqrt \beta  } \right),
\end{align}
where %$ q = \frac{{\sqrt {2m_{hh} E_0 } }}{\hbar } $,  $E_0$ is the
%barrier height,
$1\leq \chi\leq2$, $\zeta\sim 1$ are weak dimensionless functions of
$k/q$. Our further analysis will be focused on the case of the
hole's kinetic energy being substantially less then the binding
energy $E_0$, i.e. $k<<q$. For this case it is reasonable to assume
that $T_k$ does not depend on $k$. However, its rapidly decreasing
behavior for $k\gtrsim q$ has to be kept in mind as it provides
convergence for any integration over $k$ involving $T_k$. For the
case $k<<q$ the particular shape of the QW while affecting the
particular value of the tunnelimg parameter does not play an
important qualitative role and can be assumed rectangular.
 The interaction with only zeroth cylindrical harmonic means
that the continuum spectrum modified by tunneling is non-degenerate.
While this fact is not principal for the qualitative results
obtained below, it simplifies the mathematics. From the analysis
given above we conclude that the
tunneling configuration interaction exists only %to be accounted for is solemnly
between $\varphi_{k0,-\frac{3}{2}}$ and $\psi_{-\frac{3}{2}}$, and
the same interaction (governed by the parameter $T_k$ (\ref{eqT}))
is between $\varphi_{k0,+\frac{3}{2}}$ and $\psi_{+\frac{3}{2}}$.

\section{The spectrum modified by tunneling}
The transfer Hamiltonian (\ref{eqTunHamil}) with known tunneling
parameter (\ref{eqT}) allows one to construct the eigenfunctions
$\Psi$ of the whole system given those of the localized state $\psi$
and the QW states $\varphi_k$:
\begin{equation}
\label{eqPsiFano} \Psi \left( E \right) = \nu_0\left(E\right) {\psi}
+ \sum\limits_k {\nu_k \left( E \right)\varphi_k},
\end{equation}
$E$ denotes the energy of the state $\Psi$. Here $\varphi_k$ are the
wavefunctions with zeroth cylindrical harmonic, as was shown above
the other harmonics are not affected by the tunneling configuration
interaction.
 Plugging
(\ref{eqPsiFano}) into the stationary Schrodinger
 equation:
\[H\Psi=E\Psi
\] with $H$ being the effective Hamiltonian (\ref{eqTunHamil}) one
gets the following system of linear equations:
\begin{equation}
\label{eqFanosystem} \begin{array}{l}
 {\nu _0}{\varepsilon _0} + \sum\limits_k {{\nu _k}T_k^*}  = E{\nu _0}\,\,\, \\
 {\nu _k}{\varepsilon _k} + {T_k}{\nu _0} = E{\nu _k}\,\,\,\,\,\,\, \\
 \end{array}
 \end{equation}
 Solving the eignenvalue problem for system (\ref{eqFanosystem}) one
 can get the spectrum and the coefficients $\nu_0$, $\nu_k$, i.e.
the eigenfunctions of the system. Transition from discrete set of
states $\nu_k(E)$ to continuous function $\nu(\varepsilon,E)$ is
straightforward (as the continuum states are non-degenerate we can
use the energy $\varepsilon$ instead of $k$ as the quantum number of
the state). Instead of (\ref{eqPsiFano}) and (\ref{eqFanosystem}) we
write:
\begin{equation} \label{eqExpand} \Psi \left( E \right) = {\nu
_0}\left( E \right){\psi} + \int_0^\infty {\nu \left( {E,\varepsilon
} \right)} \varphi \left( \varepsilon \right)d\varepsilon,
\end{equation}
 \begin{equation}
 \label{eqFanoSystemContinous}
 \begin{array}{l}
 {\nu _0}\left(E\right){\varepsilon _0} + \int_0^\infty {t\left( \varepsilon  \right)\nu \left( E,\varepsilon  \right)d\varepsilon }  = E{\nu _0}\left(E\right),\,\,\, \\
 \nu \left( E,\varepsilon  \right)\varepsilon  + t\left( \varepsilon  \right){\nu _0}\left(E\right) = E\nu \left( E,\varepsilon  \right).\,\,\,\,\,\,\,
 \end{array}
 \end{equation}
 The normalizations for $\psi$ and
$\varphi(\varepsilon)$ are:
\begin{align} \label{eqNormUnity}
& \left\langle {{\psi}\left( {{\varepsilon _0}} \right)|{\psi}\left(
{{\varepsilon _0}} \right)} \right\rangle = 1, \nonumber
\\
& \left<\varphi(\varepsilon)|\varphi(\varepsilon')\right>= \delta
\left( {\varepsilon  - \varepsilon '} \right).
\end{align}
With the chosen normalization the discrete tunneling parameter $T_k$
and the one entering (\ref{eqFanoSystemContinous}) are related as
follows:
\begin{equation}
\label{eqtktE} T_k^2N_0\left( \varepsilon \right) = {t^2}\left(
\varepsilon \right),\end{equation} where
\begin{equation} \label{eqdensStates}N_0(\varepsilon)=%\frac{{{S^{1/2}}}}{{{\pi^{3/2}}}}
\sqrt {\frac{ m_{hh}'S}{{2\pi^3\varepsilon{\hbar
^2}}}}\end{equation} is the density of states for the selected basis
of states with zeroth cylindrical harmonic. The discrete system
(\ref{eqFanosystem}) is an eigenvalue problem, but the continuous
problem (\ref{eqFanoSystemContinous}) is not. In the present work we
consider the case of the localized energy level lying within the
range of the continuum: $\varepsilon_0>>t^2$. For this case the
solution is obtained as shown in \cite{PhysRev.124.1866}:
\begin{align}
\label{eqSolution} & {\nu _0}^2\left( E \right) = \frac{{{t^2}\left(
E \right)}}{{{\pi ^2}{t^4}\left( E \right) + {{\left( {E -
\widetilde{{\varepsilon
_0}}} \right)}^2}}}, \nonumber \\
&\nu \left( E,\varepsilon \right) = {\nu _0}\left( E \right)\left(
{P\frac{{t\left( \varepsilon \right)}}{{E - \varepsilon }} +
Z\left(E\right)t\left(E\right)}
%\frac{{E - {\varepsilon _0} - F\left(
%E \right)}}{{t\left( E \right)}}}
\delta \left( {E - \varepsilon } \right)\right),
\end{align}
 where
\begin{align}
\label{eqZ} Z\left( E \right) &= \frac{E - \varepsilon_0 - F\left(E
\right)}{t^2\left(E \right)}, \nonumber \\
F\left( E \right) &= \int_0^\infty {P\frac{{{t^2}\left( \varepsilon
\right)}}{{\left( {E - \varepsilon } \right)}}d\varepsilon},
\end{align}
 $P$ stands for the principal value and $\tilde \varepsilon_0$ is the center of configuration
 resonance, which appears to be slightly shifted from
 $\varepsilon_0$:
\begin{equation}
\label{eqResonance} \tilde \varepsilon_0(E)=\varepsilon _0 + F(E).
\end{equation}
Because of $k<<q$ it is reasonable to treat $t=$const everywhere,
except for (\ref{eqZ}) where decrease of $t$ at $E\rightarrow\infty$
is necessary for convergence of the integral. In order to analyze
the influence of the configuration interaction on the luminescence
spectra we have to calculate matrix element of operator $\hat{M}$
describing interband radiative transitions between the hybridized
hole wavefunction $\Psi(E)$ and wavefunction of an electron in the
quantum well of the conductance band $\xi_{k_el_e}$, here $k_e$ is
the magnitude of the electron wavevector, $l_e$ is the number of
cylindrical harmonic analogously to ($\ref{eqwavefuncylindr}$). We
assume that (a) there are no radiative transitions between the
localized hole wavefunction $\psi$ and the 2D electron wavefunction
$\xi_{k_el_e}$ thus the matrix element for transitions from the
localized state:
\begin{equation}
\label{eqM0} \left\langle {\xi_{k_el_e}\left| {\hat M} \right|\psi}
\right\rangle =0,\end{equation} (b) the interband radiative
transitions between the free 2D states in the QW are direct, the
matrix element given by:
\begin{equation} \label{eqMatrElement}
%M_1\left(k,l,k_e,l_e  \right)
M_0 =
  \left\langle
{\xi_{k_el_e}\left| {\hat M} \right|\varphi_{kl}} \right\rangle
=u_k\delta \left( {k - {k_e}} \right) \delta_{l,l_e},
\end{equation}
where $u_k$ is the appropriate dipole matrix element. With use of
(\ref{eqExpand}), (\ref{eqSolution}), (\ref{eqM0}) and
(\ref{eqMatrElement}) we arrive to the the matrix element for
transitions between states $\Psi(E)$ and $\xi_{k_e0}$ (according to
previous notes this matrix element differs from $M_0$ only for the
zeroth cylindrical harmonic) :
\begin{equation}
\label{eqME} M= \left\langle {\xi_{k_e0}\left| {\hat M} \right|\Psi
\left( E \right)} \right\rangle  = \nu \left( {E,\alpha {\varepsilon
_e}} \right)u\left( {\alpha {\varepsilon _e}} \right),\end{equation}
$\alpha=m_{e}/m_{hh}'$, where $m_{e}$ is the effective in-plane
electron mass, $\varepsilon_e=\hbar^2k_e^2/2m_e$. The particular
form of M (\ref{eqME}) prevents from calculation of the ratio
$M^2/M_0^2$ as done in the classical Fano resonance
calculations\cite{PhysRev.124.1866}. The latter assumes unperturbed
matrix element $M_0$ to be constant which is obviously not the case
for the direct transitions demanding the momentum conservation
(\ref{eqMatrElement}). In our case the ratio $M^2/M_0^2$ doesn't
readily give physically meaningful result due to the delta function
in (\ref{eqSolution}), one rather have to proceed to calculation of
an observable. With the Fermi's Golden Rule for the transition
probability we write:
\begin{equation} \label{eqFermi} W(\hbar \omega ) = \frac{{2\pi
}}{\hbar }\int_0^\infty\int_0^\infty {\left|M \left( {E',\varepsilon
_e }\right)\right|^2 } f_e \left( {\varepsilon _e } \right)f_h
\left( E' \right)\delta \left( {E' + \varepsilon _e  + E_g - \hbar
\omega } \right)dE'd\varepsilon _e,
\end{equation} where $E_g$ is the bandgap
and $\hbar\omega$ -- energy of radiated photon, $f_e, f_h$ -- the
energy distribution functions for the electrons and holes
respectively. To deal properly with the delta function entering
$M^2$ in (\ref{eqME}) and emerging in (\ref{eqFermi}) we pass on to
averaging the $W(\hbar\omega)$ over a small spectral interval of
width $\Omega$ centered at $\omega_0$:
\[
\widetilde{W}(\hbar\omega_0)=\frac{1}{\Omega}\int_{\omega_0-\Omega/2}^{\omega_0+\Omega/2}{W\left(\hbar\omega\right)d\omega}.
\]

Using (\ref{eqME}) and (\ref{eqSolution}) we obtain:
\begin{equation}
\label{eqRes} \widetilde{W}(\hbar \omega _0 ) = \frac{{2\pi }}{\hbar
}\frac{1}{{\hbar \Omega }}\int\limits_{\frac{{\hbar \omega _0  - E_g
- \hbar \Omega /2}}{{1 + \alpha ^{ - 1} }}}^{\frac{{\hbar \omega _0
- E_g  + \hbar \Omega /2}}{{1 + \alpha ^{ - 1} }}} \left[ {N(E') -
\frac{1}{{t^2 \left( {E'} \right)\left( {\pi ^2  + Z^2 \left( {E'}
\right)} \right)}}} \right]u^2 \left( {\alpha ^{ - 1} E'}
\right)f\left(E'\right)dE',
\end{equation}
where $f\left(E'\right)=f_e \left( {\alpha ^{ - 1} E'} \right)f_h
\left( {E'} \right)$.
%The symbol $\delta(E'-E')$ has the meaning of
%density of states:
%\begin{equation} \label{eqdeltaEEm0} \delta
%\left( {E' - E'} \right) = N(E')
%\end{equation}
The first term in brackets describes the transition rate for
radiative recombination in the QW with no account for the tunneling,
therefore $N(E')$ here is the total density of states (including not
only the zeroth but all cylindrical harmonics):
\begin{equation}
 N(E') = \frac{{m_{hh}'S}}{{2\pi {\hbar
%\delta \left( {E - E} \right)
^2}}}\end{equation} Integration assuming the functions
$\widetilde{\varepsilon}_0,t,u,f$ being constant within the range of
integration ($t(E)\equiv t$, $u(E)\equiv u$ are assumed constant
everywhere) yields :
\begin{equation} \label{eqArctan} \widetilde{W}(\hbar {\omega _0}) = \frac{{2\pi
}}{\hbar }{u^2}f\left( E \right)\left[ {\frac{{{m_{hh}'}S}}{{2\pi
{\hbar ^2}}} - \frac{1}{{\hbar \Omega \pi }}\left[
{{\rm{arctan}}\frac{{\Delta E + w}}{{\pi {t^2}}} -
{\rm{arctan}}\frac{{\Delta E - w}}{{\pi {t^2}}}} \right]} \right],
\end{equation}
 where
\begin{align}
\label{eqEomega} &E = \frac{\hbar\omega _0 - E_g}{1 + \alpha ^{-1}}
\nonumber \\
&\Delta E = E - \widetilde{{\varepsilon _0}}(E) \nonumber \\
&w  = \frac{{\hbar \Omega }}{{2\left( {1 + {\alpha ^{ - 1}}}
 \right)}}
 \end{align}
 We assume weak tunneling, $t^2$ being the smallest energy scale.
In the vicinity of resonance
\begin{equation} \label{eqWindow} \Delta E \in \left( { - w +
{t^2},w - {t^2}} \right),
 \end{equation}
expansion of (\ref{eqArctan}) to the first order in $t^2$ gives:
\begin{equation} \label{eqW}
\widetilde{W}(\hbar {\omega _0}) = \frac{{2\pi }}{\hbar
}{u^2}f\left( E \right)\left[ {\frac{{{m_{hh}'}S}}{{2\pi {\hbar
^2}}} - \frac{1}{{\hbar \Omega }} + \frac{1}{{1 + {\alpha ^{ -
1}}}}\frac{{{t^2}}}{{{w^2} - {{\left( {\Delta E} \right)}^2}}}}
\right].
\end{equation}
Note that (\ref{eqW}) has a term $-1/\hbar\Omega$ which does not
depend on the tunneling. Its appearance is due to peculiarity of the
mathematics of the Fano model reflected in (\ref{eqSolution}). When
a non-interacting state with energy $\varepsilon_0$ is appended to
the system so that $\varepsilon_0$ lies within its spectrum, one of
the energy levels of the whole system becomes doubly degenerate.
This fact is not properly accounted for in (\ref{eqSolution}) and
one state is lost.
 It should be added back manually to the spectral density by canceling
the second term in (\ref{eqW}). Treating the same issue in a
different way, one should examine
% \label{eqDeltaW}
$\Delta\widetilde{W}  = \widetilde{W} - \widetilde{W_0}$
%\end{equation}
instead of $W$ itself, $\widetilde{W_0}$ being the unperturbed
transition rate: (\ref{eqArctan}) evaluated for $t=0$. In a similar
way studying the ratio of matrix elements in original Fano work
\cite{PhysRev.124.1866} circumvents the disappearance of one level.

The results obtained for a single impurity can be also applied to an
ensemble of impurities provided their interaction between each other
is weak compared to the tunnel coupling with the QW. If the
concentration of the impurities is low enough to produce only weak
perturbation of the luminescence spectra, we can simply multiply the
tunneling term by the number of impurities. After normalization by
the area of the QW we finally get the spectral density of the
luminescence intensity:
\begin{equation}
\label{eqArcTanSgn}I(\hbar {\omega _0}) = \frac{{2\pi }}{\hbar
}{u^2}f\left( E \right)\left[ {\frac{{{m_{hh}'}}}{{2\pi {\hbar ^2}}}
+ \frac{n}{{\pi\hbar \Omega  }}\left[ {{\rm{arctan}}\frac{{\Delta E
- w}}{{\pi {t^2}}} - {\rm{arctan}}\frac{{\Delta E + w}}{{\pi {t^2}}}
- \pi \frac{{{\mathop{\rm sgn}} \left( {\Delta E - w} \right) -
{\mathop{\rm sgn}} \left( {\Delta E + w} \right)}}{2}} \right]}
\right],
\end{equation}
where $n$ is the 2D concentration of impurities. The last term in
brackets corrects the lost level issue to provide exact canceling of
the perturbation of the spectra at $t=0$. For high concentration of
the impurities the formula (\ref{eqArcTanSgn}) may give a
meaningless result (the intensity may become negative at some
points). Indeed for high concentration the real physical picture
becomes slightly different -- interaction between the impurities
splits their energy levels forming a small range of discrete levels,
accordingly, the configuration resonances become slightly shifted.
Taking this effect into account eliminates the puzzling behavior of
(\ref{eqArcTanSgn}) at high concentration but does not affect the
answer for the calculation of polarization given in the next
section.

\begin{figure}
  \leavevmode
 \centering\includegraphics[width=0.5\textwidth]{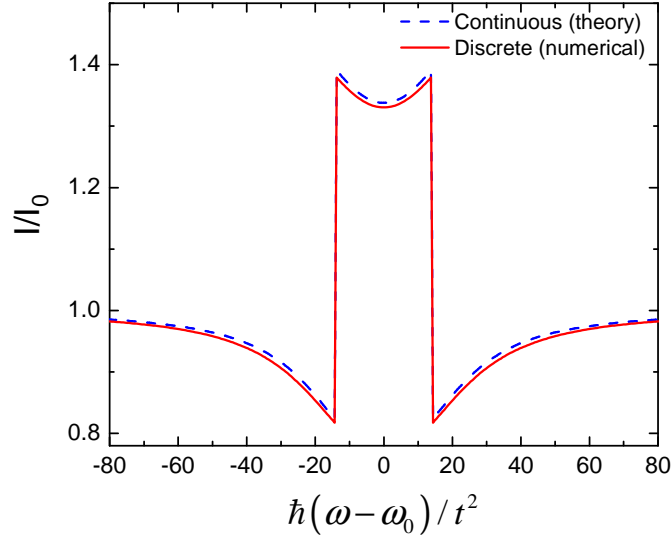}
 \caption{(Color online) Modification of the luminescence spectrum
 by tunneling configuration interaction: numerical calculation for the discrete levels (solid line) and analytical
 formula (\ref{eqArcTanSgn}) (dashed line)}
 \label{fig_Enhancement}
\end{figure}

The analytical result (\ref{eqArcTanSgn}) was verified by numerical
simulation performed for the discrete system (\ref{eqFanosystem}).
The system was solved for 500 discrete levels with interlevel
separation $10^{-5}$ eV, the discrete tunneling parameter was taken
$T_k=3.3\cdot10^{-5}$ eV, which corresponds to the continuous
tunneling parameter $t^2=10^{-4}$ eV, the other relevant parameters
were: $w=5\cdot10^{-4}$ eV, $n=10^{10}$ cm$^{-2}$, $m_e=0.03$ $m_0$,
$m_{hh}=0.5$ $m_0$, $m_{hh}'=0.15$ $m_0$. In both calculations all
the states were assumed fully occupied, i.e. the energy distribution
function was kept $f(E)=1$.
 Analogously to (\ref{eqMatrElement}) the matrix
element for the discrete system was taken: ${M_k}\left( {\varepsilon
,{\varepsilon _e}} \right) = {u_k}{\delta _{k,{k_e}}}\delta
_{l,{l_e}}$. The calculation result presented in
Fig.\ref{fig_Enhancement} demonstrates perfect agreement with the
analytical expression (\ref{eqArcTanSgn}) and confirms the validity
of the latter.
The considered tunneling configuration interaction thus gives rise
to the luminescence intensity within a certain spectral range
(\ref{eqWindow}) corresponding to the resonance. This increase is
compensated by the decrease of the intensity outside of the this
range as can be seen in Fig.\ref{fig_Enhancement}. The width of the
resonance is determined by $\Omega$ which has the meaning of
spectral resolution of the measurement setup. However, for
comparison with experimental spectra the inhomogeneous broadening
should be accounted for as it usually exceeds the instrumental
spectral resolution.  An expression for the integral intensity over
the whole spectra reads:
\begin{equation}
\label{eqIntegral} I = \frac{{2\pi }}{\hbar }\int\limits_0^\infty
{{u^2}f\left( E \right)\left[ {\frac{{{m_{hh}'}}}{{2\pi {\hbar ^2}}}
- \frac{{n{t^2}\left( E \right)}}{{{\pi ^2}{t^4}\left( E \right) +
{{\left( {E - \widetilde{{\varepsilon _0}}} \right)}^2}}} + n\delta
\left( {E - \widetilde{{\varepsilon _0}}} \right)} \right]dE}.
\end{equation}
This formula follows from (\ref{eqFermi}) in the same way as
(\ref{eqArctan}) and (\ref{eqW}) were obtained. The delta-function
here is added manually to treat the lost level issue  -- it provides
canceling of the second term in the limit $t\rightarrow 0$ and thus
gives the correct expression in the absence of tunneling:
\begin{equation}
\label{eqI0} {I_0} = \frac{{2\pi }}{\hbar }\int\limits_0^\infty
{{u^2}f\left( E \right)\frac{{{m_{hh}'}}}{{2\pi {\hbar ^2}}}dE}.
\end{equation} Note that the spectral width of the resonance
$\Omega$ does not enter the expression for integral intensity
(\ref{eqIntegral}).

\section{Polarization of the spectra}
\label{SecPolar} 
\begin{figure}
 \label{figscheme}
  \leavevmode
 \centering\includegraphics[width=0.4\textwidth]{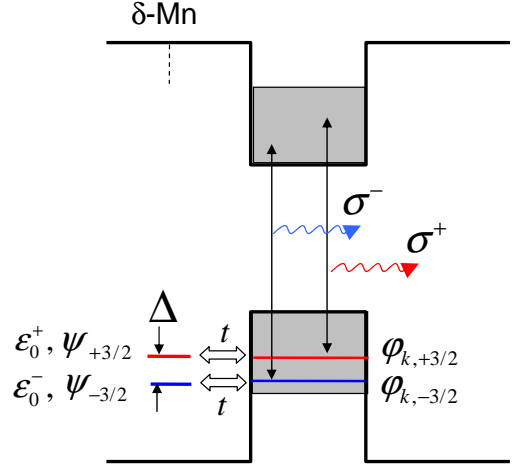}
 \caption{(Color online) Mechanism of polarization of the luminescence. The localized hole levels split
 in magnetic field. Each of them effectively couples with
 the 2D holes having certain projection of angular momentum. Shifted positions of the resonances
 with account for temperature distribution of the holes cause the difference
 in intensities of circular polarizations $\sigma^+$, $\sigma^-$}
 \label{fig_scheme}
\end{figure}

While the redistribution of the spectral density
does not change the integral intensity it gives rise to the integral
polarization of the spectra in the magnetic field as illustrated by
Fig.\ref{fig_scheme}. The 2D holes with projection of total angular
momentum $j=+3/2$ and $j=-3/2$ recombine emitting respectively
right- ($\sigma^+$) and left- ($\sigma^-$) circularly polarized
light. In section \ref{secT} it was shown that the heavy holes with
$j=-3/2$ interact basically with the eigenfunction
$\psi_{-\frac{3}{2}}$ of the localized state. Let us denote the
corresponding energy of this state $\varepsilon_0^-$. The 2D holes
with $j=+3/2$ interact in turn with $\psi_{+\frac{3}{2}}$ which
corresponds to the energy $\varepsilon_0^+$. An external magnetic
field applied in $z$ would cause Zeeman splitting between
$\varepsilon_0^+$ and $\varepsilon_0^-$. The splitting
$\Delta=\varepsilon_0^+-\varepsilon_0^-$ may also originate from
exchange interaction of holes with spin-polarized Mn ions. The value
of $\Delta$ in this case is determined by exchange constant and
depend on the degree of Mn spin polarization. The splitting of the
localized energy level leads, in turn, to the splitting of the
configuration resonance. Indeed, as follows from (\ref{eqRes}),
(\ref{eqArctan}),(\ref{eqZ}) the difference in the positions of the
resonances $E_+$ and $E^-$ corresponding to the localized levels
$\varepsilon_0^+$ and $\varepsilon_0^-$ is given by:
\begin{equation}\label{eqdeltatilde}\widetilde\Delta  = {E^ + } - {E^ - } = \Delta  + {t^2}\ln
\left( {1 + \frac{{\widetilde\Delta }}{{{E_ - }}}}
\right).\end{equation} Unless the position of the resonance $E^-$ is
too close to the valence band edge the last term in
(\ref{eqdeltatilde}) can be neglected and
$\widetilde{\Delta}=\Delta= \varepsilon_0^+-\varepsilon_0^-$. The
applicability of this result is limited to the case
$\varepsilon_0>\Delta$. This condition, in fact, simply means that
the splitting of the localized level does not bring any of the
sublevels beyond the energy range of the 2D continuum so that the
exploited Fano approach remains valid. Our consideration will be
always limited to this case.
With account of the energy distribution functions for the holes and
electrons the shifted positions of the resonances lead to the
difference in the luminescence intensity for the opposite circular
polarizations. Let $I^+$, $I^-$ be the integral luminescence
intensities of circular polarizations $\sigma^+$  and $\sigma^-$
respectively. Assuming $\left|I^\pm-I_0\right|<<I_0$ the integral
polarization is given by:
\[
P = \frac{{I^ +   - I^ -  }}{{2I_0 }}.
\]
With use of (\ref{eqIntegral}) this yields: \begin{equation}
\label{eqPolInt} P = \frac{{n\pi {\hbar ^2}}}{{{m_{h{h^\prime
}}}}}\frac{{f\left( {{}\widetilde{\varepsilon} _0^ - } \right) -
f\left( {{}\widetilde {\varepsilon} _0^ + } \right) +
\int\limits_0^\infty {\left[ {\frac{{{t^2}\left( E \right)f\left( E
\right)}}{{{\pi ^2}{t^4}\left( E \right) + {{\left( {E - {\rm{
}}\widetilde{\varepsilon} _0^ + } \right)}^2}}} - \frac{{{t^2}\left(
E \right)f\left( E \right)}}{{{\pi ^2}{t^4}\left( E \right) +
{{\left( {E - {\rm{ }}\widetilde{\varepsilon} _0^ - } \right)}^2}}}}
\right]dE} }}{{\int\limits_0^\infty  {f\left( E \right)dE}
}}.\end{equation}
 The slow varying functions in the upper integrals may be assumed
as constants taken at $\widetilde{\varepsilon}_0^-,
\widetilde{\varepsilon}_0^+$, the tunneling parameter will be
treated as a constant in the whole range of interest $t^2(E)\equiv
t^2$.

\begin{figure}
 \label{fignondegen}
  \leavevmode
 \centering\includegraphics[width=0.5\textwidth]{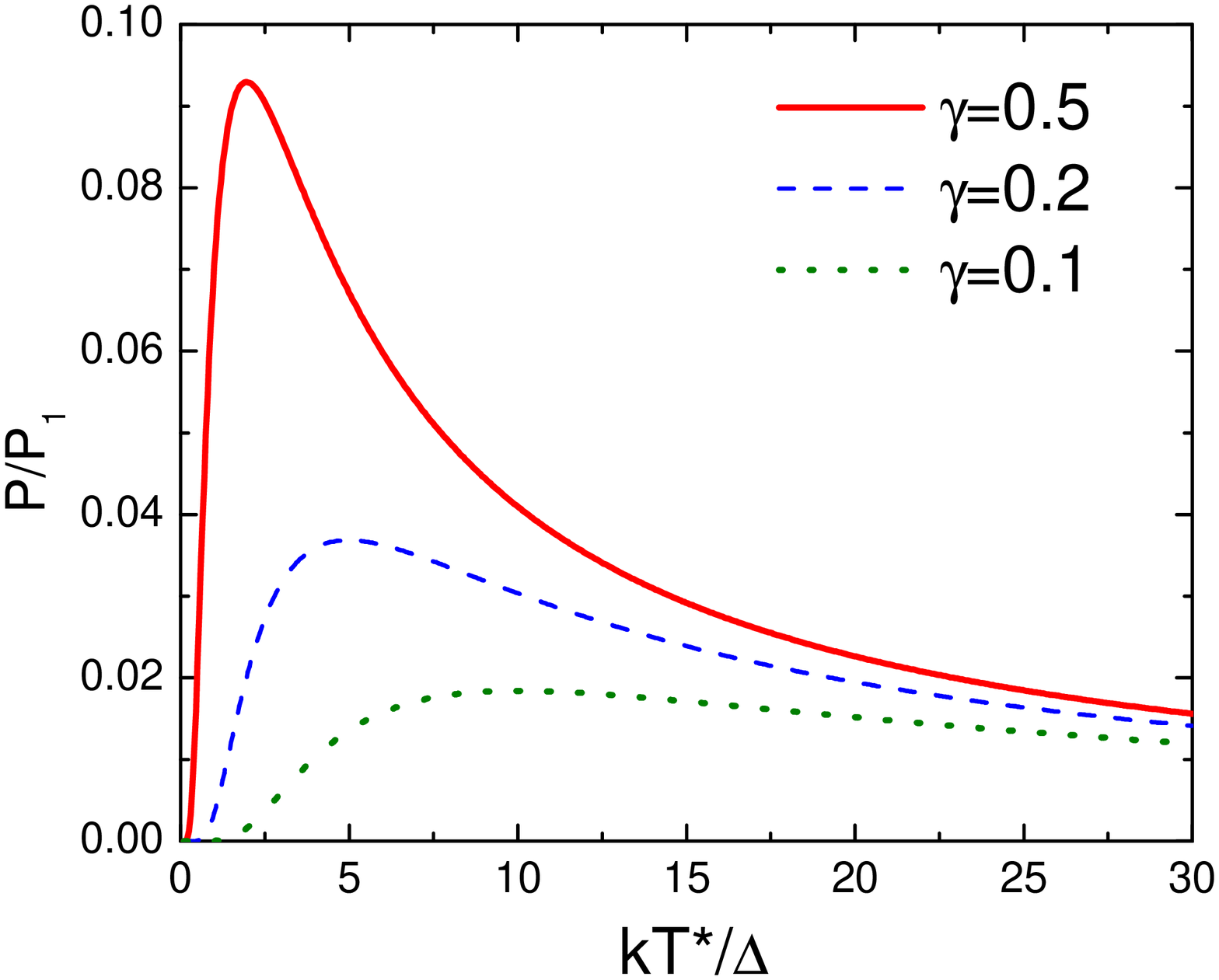}
 \caption{(Color online) Temperature dependence of integral polarization.
Electrons are non-degenerate, holes are either non-degenerate or
have the constant distribution function for different values of
parameter $\gamma \equiv \Delta /\widetilde{\varepsilon _0 }$. }
  \end{figure}

\begin{figure}
 \label{figDgenerate}
  \leavevmode
 \centering\includegraphics[width=0.5\textwidth]{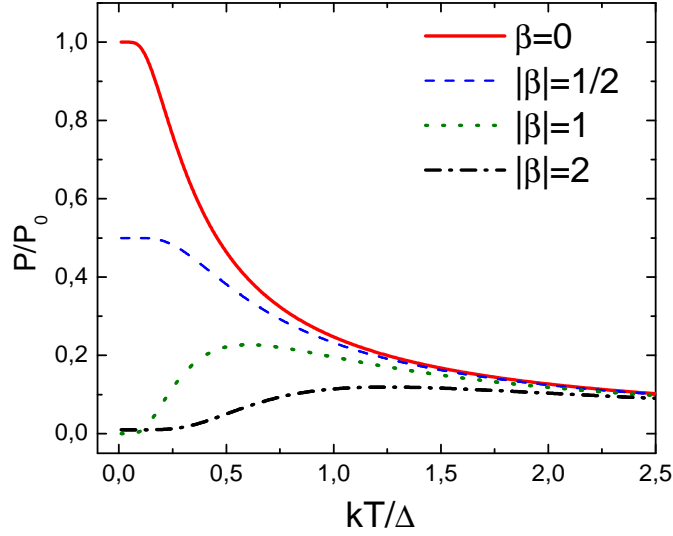}
 \caption{(Color online) Temperature dependence of polarization for the case of electrons distribution function being constant
 within the configuration resonances. The parameter
 $\beta  \equiv \frac{\widetilde{\varepsilon _0} - \mu _h }{\Delta}$
 denotes deviation of the holes Fermi level from the configuration
 resonance, $\gamma=0.1$.}
\end{figure}

Then expanding over $t^2$ gives for the first order term:
\begin{equation}
\label{eqPolGeneral} P = \frac{{n\pi {\hbar
^2}{t^2}}}{{{m_{hh}'}}}\frac{{{f}\left( {\widetilde{\varepsilon} _0^
- } \right){{\left( {\widetilde{\varepsilon} _0^ - } \right)}^{ -
1}} - {f}\left( {\widetilde{\varepsilon} _0^ + } \right){{\left(
{\widetilde{\varepsilon} _0^ + } \right)}^{ -
1}}}}{{\int\limits_0^\infty  {{f}\left( E \right)dE} }}.
\end{equation}
The formula (\ref{eqPolGeneral}) leaves not much room for further
simplification for a general case of $f_e(E), f_h(E)$ being two
Fermi distributions characterized by chemical potentials $\mu_e$,
$\mu_h$ and the temperatures $T_e$ and $T_h$ respectively, all four
parameters being different. Let us analyze a few particular cases
leading to compact analytical expressions for P. All the cases imply
$\widetilde{\varepsilon_0}>\Delta$. Firstly, let the holes be fully
degenerate and both energies $\widetilde{\varepsilon}_0^-$ and
$\widetilde{\varepsilon_0}^+$ lying well beyond the quasi Fermi
level of the holes so that: ${\mu _h  - \widetilde{\varepsilon} _0^+
}
>  > \Delta$. In this case the distribution function of the holes can be considered
as constant $F_h(E)=1$ in the range $E \in \left(
{\widetilde{\varepsilon} _0^ - ,\widetilde{\varepsilon} _0^ +  }
\right)$ Assuming the electrons to be non-degenerate with their
temperature $T_e$ the formula (\ref{eqPolGeneral}) reduces to:
\begin{equation}
\label{eqNonDeg}P = P_1 {e^{ - \frac{{\widetilde{{\varepsilon
_0}}}}{{k{T^*}}}}}\sinh \frac{\Delta }{{2k{T^*}}},\end{equation}
where
\[
P_1=\frac{{2\pi n{\hbar ^2}{t^2}}}{{{m_{hh}'}\widetilde{{\varepsilon
_0}}k{T^*}}}.
\]
Here $T^*=\alpha T_e$. Exactly the same expression is valid for the
case when both electrons and holes are non-degenerate. The only
difference from the previously considered case is that now the
effective temperature $T^*$ is given by
\[\frac{1}{{{T^*}}} = \left( {\frac{1}{{\alpha {T_e}}} + \frac{1}{T}} \right).\]
The expression (\ref{eqNonDeg}) is plotted in Fig.\ref{fignondegen}
for different values of the parameter $\gamma  \equiv \Delta
/\widetilde{\varepsilon _0 }$. The polarization shows nonmonotonous
behavior with increasing the temperature. In the discussed theory
the polarization arises from splitting of the configuration
resonances positions for $\sigma^+$ and $\sigma^-$ spectra. The
configuration resonance itself causes the redistribution of the
transitions rate in the vicinity of the resonance energy conserving
the total rate, thus the net polarization is subect to the
difference in occupation numbers for $\widetilde{\varepsilon}_0^-$
and $\widetilde{\varepsilon _0}^+$. The maximum integral
polarization is therefore naturally expected when the derivative of
the combined distribution function $f(E)$ reaches its maximum value
within the range $E \in \left( {\widetilde{\varepsilon} _0^ -
,\widetilde{\varepsilon} _0^ +  } \right)$. For the considered case
the maximum of the derivative is at ${\widetilde{\varepsilon} _0}$
when ${\widetilde{\varepsilon} _0} = k{T^*}$ and the value of the
derivative decreases with increase of $\widetilde{\varepsilon}_0$.
This explains the overall decrease of the maximum polarization with
decrease of $\gamma$ in Fig.\ref{fignondegen}.
For another case we consider the electrons distribution function
$f_e$ being nearly constant within the configuration resonances.
This can be due to their non-equilibrium distribution with a high
quasi Fermi level or the electrons temperature $T_e$ being much
higher than that of the holes. The holes are now considered to have
Fermi distribution function with the quasu Fermi level $\mu_h$ and
the temperature $T$. We also assume
%$\Delta<<\widetilde{\varepsilon}_0$,
$kT<<\widetilde{\varepsilon}_0$. In this case from
(\ref{eqPolGeneral}) we get:
%The simplified expression reads:
\begin{equation}
\label{eqPolarDimensionless} P =P_0 \left( {\frac{{2\exp \left(
\beta\xi \right)\sinh \left( {\xi /2} \right) + \gamma }}{{\exp
\left( {2\beta\xi } \right) + 2\exp \left( \beta\xi \right)\cosh
\left( {\xi /2} \right) + 1}}} \right),
\end{equation}
where
\begin{align}
& \beta = \frac{{\widetilde{\varepsilon _0 } - \mu _h }}{{\Delta}}, \nonumber \\
%& \eta  = \widetilde{\varepsilon _0 }/kT \nonumber \\
&  \xi  = \Delta/kT \nonumber, \\
& P_0  = \frac{{n\pi \hbar ^2 t^2 }}{{m_{hh}' \mu _h ^2 }}.
 \end{align}
 The dependence (\ref{eqPolarDimensionless}) of $P/P_0$ on $1/\xi$ is plotted in
Fig.\ref{figDgenerate} for different values of the parameter $\beta$
(the value of $\gamma$ was taken $0.1$). In this case the maximum of
the distribution function derivative is at the holes Fermi level
$\mu_h$, therefore the largest integral polarization corresponds to
$\beta=0$.
For this particular case (\ref{eqPolarDimensionless}) simplifies
into:
\begin{equation}
\label{eqTanh} P = P_0\left( {\tanh \left( \xi  \right) +
\frac{\gamma }{{2\cosh ^2 \left( \xi \right)}}} \right).
\end{equation}
The integral polarization obtained within given approach can be
alternatively expressed through an effective g-factor of the holes
$g_{eff}$. Let us consider Zeeman term in the Hamiltonian of the 2D
holes:
\[H_B=\mu_0 g_{eff}J_zB,\] where $J_z$ is the angular momentum
projection operator, $\mu_0$ is the Bohr magneton and B is the
magnetic field applied along $z$-- direction. The polarization of 2D
holes due to the splitting between the two subbands with $j=+3/2$
and $j=-3/2$ is given by: \begin{equation}{P_B} =
\frac{{\int\limits_0^\infty {\left( {f\left( {E - {\Delta_B} /2}
\right) - f\left( {E + \Delta_B /2} \right)} \right)dE}
}}{{\int\limits_0^\infty  {\left( {f\left( {E - \Delta_B /2} \right)
+ f\left( {E + \Delta_B /2} \right)} \right)dE} }},\end{equation}
where $\Delta_B=3\mu_0 g_{eff}B$. For the nondegenerate case one
gets
\begin{equation}\label{eqPbnondeg}P_B = \tanh
\frac{\Delta_B }{{2kT^*}}\end{equation} Comparing (\ref{eqPbnondeg})
with (\ref{eqNonDeg}) gives:
\begin{equation}
{g_{eff}} = \frac{{2k{T^*}}}{{3{\mu _0}B}}{\tanh ^{ - 1}}\left[
{\left( {\frac{{2\pi n{\hbar ^2}{t^2}}}{{{m_{h{h^\prime
}}}\widetilde{{\varepsilon _0}}kT^*}}} \right){e^{ -
\frac{{\widetilde{{\varepsilon _0}}}}{{kT^*}}}}\sinh \frac{{{\Delta
_0}}}{{2kT^*}}} \right]
%\widetilde\Delta  = 2k{T^*}{\tanh ^{ - 1}}\left[ {\left(
%{\frac{{2\pi n{\hbar ^2}{t^2}}}{{{m_{h{h^\prime
%}}}\widetilde{{\varepsilon _0}}kT}}} \right){e^{ -
%\frac{{\widetilde{{\varepsilon _0}}}}{{kT}}}}\sinh \frac{{{\Delta
%_0}}}{{2kT}}} \right]
\end{equation}
In the same way an expression for the degenerate case can be easily
obtained.

\section{Discussion}
\label{secDiscus}
\begin{figure}
 \label{figSpectra}
  \leavevmode
 \centering\includegraphics[width=0.5\textwidth]{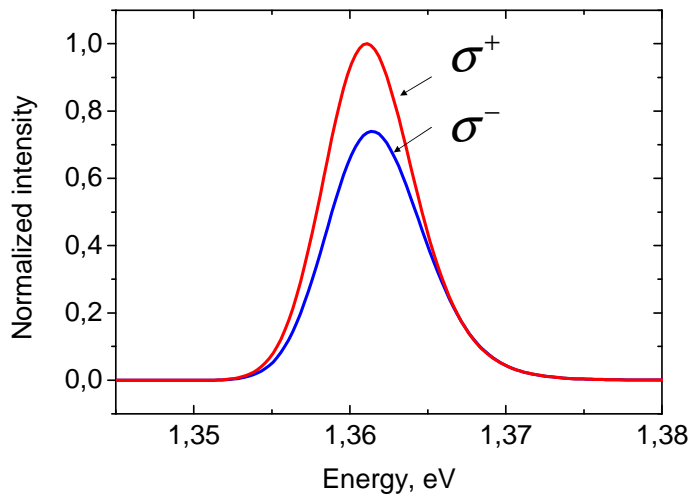}
 \caption{(Color online) An example of calculated luminescence spectra for right ($\sigma^+$) and left ($\sigma^-$) circular polarizations.
  The parameters used in calculations are given in the text. }
\end{figure}

%\section{Estimations for the values}
The key advantage of the Fano approach utilized in the present study
is that the unknown eigenfunctions of the complex system are
expressed through the known ones of the uncoupled states, in our
case these are the hole localized at Mn and the free 2D hole in the
QW. Given the expansion (\ref{eqPsiFano}) any effects on the
localized state can be easily translated into effect for the whole
coupled system. For the polarization effect under study the key
parameters are the energy of the localized level $\varepsilon_0$,
the splitting parameter $\Delta$ and the tunneling parameter $t^2$.
The binding energy for a hole at a single Mn in GaAs is known to be
$E_0\approx110$ meV \cite{PhysRevB.10.2501}. For the enhanced Mn
concentrations in the delta-layer up to $10^{13}$ cm$^{-2}$ the
impurity band is established with the binding energy lowering down
to 50 meV or even less
\cite{PhysRevB.73.235208,RevModPhys.78.809,PhysRevB.8.3803}. Thus we
consider the QW depth for the holes starting from 50 meV to be
likely for the resonance tunneling effects under study to occur.
Estimations for the splitting energy $\Delta$ subject to both
exchange interaction between the hole and Mn and external magnetic
field. For small concentration of Mn the splitting between the
states having opposite projections of the angular momentum can be
simply estimated as $\Delta=\mu_0gB$ with $B$ being an external
magnetic field and $g\approx3$ is the g-factor for the hole at Mn.
This makes $\Delta\sim0.1$ meV for $B\sim 1$ T. Samples with higher
Mn concentrations up to a few percent are known to exhibit
ferromagnetic properties \cite{RevModPhys.78.809}, in this case the
levels splitting $\Delta$ is to be considered with account for p-d
exchange interaction \cite{RevModPhys.78.809}. The particular value
of $\Delta$ for typical experimental samples still remains
questionable, in the ferromagnetic regime the splitting is probably
believed to be in the range $\Delta\sim 1 - 10$ meV by the order of
magnitude. The magnitude of the tunnel coupling is, of course, the
key parameter determining the polarization. From (\ref{eqT}) and
(\ref{eqdensStates}) it follows that the tunneling parameter can be
estimated as
\begin{equation}
\label{eqt2} t^2  = \left( {A - B} \right)^2 \frac{{2m_{hh}m_{hh}'
 E_0 }}{{\hbar ^4 }}e^{ - \frac{{2 \sqrt {2m_{lh} E_0 }
}}{\hbar } d}.
\end{equation}
Substituting band parameters for GaAs $B-A\approx
1\cdot\hbar^2/2m_0$, $E_0=100$ meV, $d=4$ nm one obtains the
characteristic value $t^2 \sim 0.1$ meV. We then take the Mn
concentration $n\sim10^{12}$ cm$^{-2}$, $\xi=1$,
$\varepsilon_0\sim3$ meV, $\beta=0$. This set of parameters gives
$P_0\approx0.2$, $P_1\approx0.6$. The experimental temperature
dependence of polarization obtained in
\cite{springerlink:10.1134/S0021364009220056} qualitatively agrees
with (\ref{eqTanh}). Beside the analytical expressions for the
general case (\ref{eqPolGeneral}) and particular cases
(\ref{eqNonDeg}),(\ref{eqTanh}), a numerical simulation of the
luminescence spectra can be performed based on (\ref{eqArcTanSgn}).
An example of such calculation is shown in Fig.\ref{figSpectra}. For
the calculation the following parameters were taken: $\Delta=1$ meV,
$n=10^{11}$ cm$^{-2}$,  $T=T_{e}=20$ K, $\varepsilon_0=\mu_{h}=1$
meV, the inhomogeneous broadening of the spectra was accounted for
by normal distribution of $E_g$ with dispersion taken as $\sigma=3$
meV (corresponds to the fluctuation of the QW width by half a
monolayer). The calculated spectra presented in Fig.\ref{figSpectra}
seem to be in good agreement with the experimental results obtained
in
\cite{springerlink:10.1134/S0021364009220056,springerlink:10.1134/S1063783410110144}.

As mentioned in section \ref{secStates} the magnetic field in the
developed theory was assumed nonquantizing. Indeed an estimate for
the energy of Landau levels separation gives:
\[\hbar\omega_c=\frac{e\hbar B}{m_{hh}'c}\approx 0.3\,\rm{meV}\] for
the magnetic field $B=0.5$ T. This value is substantially less than
the typical kinetic energy of the holes estimated as
$\varepsilon\approx 1-10$ meV. However, this value is comparable
with the tunneling parameter $t^2$. Therefore for the experimental
data the validity of the developed theory is well justified for
$B\lesssim0.5$ T.

\section{Summary}
We have presented a theory treating tunneling configuration
interaction between a continuum of states in the QW and a
paramagnetic impurity located outside of the QW. The coupling
between the localized state and the QW is described by means of
tunneling Hamiltonian. We utilized the well known Fano approach for
calculation of the matrix elements for direct radiative transitions
between electrons and holes in the QW. At that the new results were
obtained not covered by the conventional Fano formula
\cite{PhysRev.124.1866,RevModPhys.82.2257}, the tunneling
interaction between the localized hole state and continuum of states
in the QW results in the symmetrical redistribution of the
luminescence spectral density in the vicinity of the resonance.
While this redistribution does not affect integral luminescence
intensity, it causes an integral circular polarization of the light
emitted from the QW provided the localized state is split in the
projection of the hole angular momentum under external magnetic
field or due to exchange interaction with Mn ions. The presented
theory expresses the eigen states of the system with weak tunnel
coupling through the wavefunctions of the hole localized at
paramagnetic center and the 2D states of the continuum. For this
reason it seems to be capable of describing other effects expected
in such systems like an anisotropy of the holes g-factor in the QW
induced by a paramagnetic impurity  or the indirect exchange
interaction between the localized hole states provided by the
continuum located at a tunneling distance.
\section{acknowledgements}
We thank V. D. Kulakovskii for very fruitful discussions and also
express our thanks to B. A. Aronzon, P. I. Arseev, V. L. Korenev, V.
F. Sapega for useful and helpful comments. The work has been
supported by RFBR (grants no 09-02-00469, 11-02-00348, 11-02-00146,
12-02-00815,12-02-00141), Russian Ministry of Education and Science
(contract N 14.740.11.0892).
\bibliography{MnFano}

\begin{thebibliography}{17}
\expandafter\ifx\csname natexlab\endcsname\relax\def\natexlab#1{#1}\fi
\expandafter\ifx\csname bibnamefont\endcsname\relax
  \def\bibnamefont#1{#1}\fi
\expandafter\ifx\csname bibfnamefont\endcsname\relax
  \def\bibfnamefont#1{#1}\fi
\expandafter\ifx\csname citenamefont\endcsname\relax
  \def\citenamefont#1{#1}\fi
\expandafter\ifx\csname url\endcsname\relax
  \def\url#1{\texttt{#1}}\fi
\expandafter\ifx\csname urlprefix\endcsname\relax\def\urlprefix{URL }\fi
\providecommand{\bibinfo}[2]{#2}
\providecommand{\eprint}[2][]{\url{#2}}

\bibitem[{\citenamefont{Fano}(1961)}]{PhysRev.124.1866}
\bibinfo{author}{\bibfnamefont{U.}~\bibnamefont{Fano}}, \bibinfo{journal}{Phys.
  Rev.} \textbf{\bibinfo{volume}{124}}, \bibinfo{pages}{1866}
  (\bibinfo{year}{1961}).

\bibitem[{\citenamefont{Miroshnichenko
  et~al.}(2010)\citenamefont{Miroshnichenko, Flach, and
  Kivshar}}]{RevModPhys.82.2257}
\bibinfo{author}{\bibfnamefont{A.~E.} \bibnamefont{Miroshnichenko}},
  \bibinfo{author}{\bibfnamefont{S.}~\bibnamefont{Flach}}, \bibnamefont{and}
  \bibinfo{author}{\bibfnamefont{Y.~S.} \bibnamefont{Kivshar}},
  \bibinfo{journal}{Rev. Mod. Phys.} \textbf{\bibinfo{volume}{82}},
  \bibinfo{pages}{2257} (\bibinfo{year}{2010}).

\bibitem[{\citenamefont{Blom et~al.}(2002)\citenamefont{Blom, Odnoblyudov,
  Yassievich, and Chao}}]{PhysRevB.65.155302}
\bibinfo{author}{\bibfnamefont{A.}~\bibnamefont{Blom}},
  \bibinfo{author}{\bibfnamefont{M.~A.} \bibnamefont{Odnoblyudov}},
  \bibinfo{author}{\bibfnamefont{I.~N.} \bibnamefont{Yassievich}},
  \bibnamefont{and} \bibinfo{author}{\bibfnamefont{K.~A.} \bibnamefont{Chao}},
  \bibinfo{journal}{Phys. Rev. B} \textbf{\bibinfo{volume}{65}},
  \bibinfo{pages}{155302} (\bibinfo{year}{2002}).

\bibitem[{\citenamefont{Okulov et~al.}(2011)\citenamefont{Okulov, Lonchakov,
  Govorkova, Okulova, Podgornykh, Paranchich, and Paranchich}}]{okulov:220}
\bibinfo{author}{\bibfnamefont{V.~I.} \bibnamefont{Okulov}},
  \bibinfo{author}{\bibfnamefont{A.~T.} \bibnamefont{Lonchakov}},
  \bibinfo{author}{\bibfnamefont{T.~E.} \bibnamefont{Govorkova}},
  \bibinfo{author}{\bibfnamefont{K.~A.} \bibnamefont{Okulova}},
  \bibinfo{author}{\bibfnamefont{S.~M.} \bibnamefont{Podgornykh}},
  \bibinfo{author}{\bibfnamefont{L.~D.} \bibnamefont{Paranchich}},
  \bibnamefont{and} \bibinfo{author}{\bibfnamefont{S.~Y.}
  \bibnamefont{Paranchich}}, \bibinfo{journal}{Low Temperature Physics}
  \textbf{\bibinfo{volume}{37}}, \bibinfo{pages}{220} (\bibinfo{year}{2011}),
  \urlprefix\url{http://link.aip.org/link/?LTP/37/220/1}.

\bibitem[{\citenamefont{Aleshkin et~al.}(2008)\citenamefont{Aleshkin,
  Gavrilenko, Odnoblyudov, and
  Yassievich}}]{springerlink:10.1134/S1063782608080034}
\bibinfo{author}{\bibfnamefont{V.}~\bibnamefont{Aleshkin}},
  \bibinfo{author}{\bibfnamefont{L.}~\bibnamefont{Gavrilenko}},
  \bibinfo{author}{\bibfnamefont{M.}~\bibnamefont{Odnoblyudov}},
  \bibnamefont{and}
  \bibinfo{author}{\bibfnamefont{I.}~\bibnamefont{Yassievich}},
  \bibinfo{journal}{Semiconductors} \textbf{\bibinfo{volume}{42}},
  \bibinfo{pages}{880} (\bibinfo{year}{2008}), ISSN \bibinfo{issn}{1063-7826},
  \bibinfo{note}{10.1134/S1063782608080034},
  \urlprefix\url{http://dx.doi.org/10.1134/S1063782608080034}.

\bibitem[{\citenamefont{Korenev}(2003)}]{springerlink:10.1134/1.1641485}
\bibinfo{author}{\bibfnamefont{V.}~\bibnamefont{Korenev}},
  \bibinfo{journal}{JETP Letters} \textbf{\bibinfo{volume}{78}},
  \bibinfo{pages}{564} (\bibinfo{year}{2003}), ISSN \bibinfo{issn}{0021-3640},
  \bibinfo{note}{10.1134/1.1641485},
  \urlprefix\url{http://dx.doi.org/10.1134/1.1641485}.

\bibitem[{\citenamefont{Aronzon et~al.}(2008)\citenamefont{Aronzon, Kovalchuk,
  Pashaev, Chuev, Kvardakov, Subbotin, Rylkov, Pankov, Likhachev, Zvonkov
  et~al.}}]{0953-8984-20-14-145207}
\bibinfo{author}{\bibfnamefont{B.~A.} \bibnamefont{Aronzon}},
  \bibinfo{author}{\bibfnamefont{M.~V.} \bibnamefont{Kovalchuk}},
  \bibinfo{author}{\bibfnamefont{E.~M.} \bibnamefont{Pashaev}},
  \bibinfo{author}{\bibfnamefont{M.~A.} \bibnamefont{Chuev}},
  \bibinfo{author}{\bibfnamefont{V.~V.} \bibnamefont{Kvardakov}},
  \bibinfo{author}{\bibfnamefont{I.~A.} \bibnamefont{Subbotin}},
  \bibinfo{author}{\bibfnamefont{V.~V.} \bibnamefont{Rylkov}},
  \bibinfo{author}{\bibfnamefont{M.~A.} \bibnamefont{Pankov}},
  \bibinfo{author}{\bibfnamefont{I.~A.} \bibnamefont{Likhachev}},
  \bibinfo{author}{\bibfnamefont{B.~N.} \bibnamefont{Zvonkov}},
  \bibnamefont{et~al.}, \bibinfo{journal}{Journal of Physics: Condensed Matter}
  \textbf{\bibinfo{volume}{20}}, \bibinfo{pages}{145207}
  (\bibinfo{year}{2008}),
  \urlprefix\url{http://stacks.iop.org/0953-8984/20/i=14/a=145207}.

\bibitem[{\citenamefont{Aronzon et~al.}(2010)\citenamefont{Aronzon, Pankov,
  Rylkov, Meilikhov, Lagutin, Pashaev, Chuev, Kvardakov, Likhachev, Vihrova
  et~al.}}]{10.1063/1.3267314}
\bibinfo{author}{\bibfnamefont{B.~A.} \bibnamefont{Aronzon}},
  \bibinfo{author}{\bibfnamefont{M.~A.} \bibnamefont{Pankov}},
  \bibinfo{author}{\bibfnamefont{V.~V.} \bibnamefont{Rylkov}},
  \bibinfo{author}{\bibfnamefont{E.~Z.} \bibnamefont{Meilikhov}},
  \bibinfo{author}{\bibfnamefont{A.~S.} \bibnamefont{Lagutin}},
  \bibinfo{author}{\bibfnamefont{E.~M.} \bibnamefont{Pashaev}},
  \bibinfo{author}{\bibfnamefont{M.~A.} \bibnamefont{Chuev}},
  \bibinfo{author}{\bibfnamefont{V.~V.} \bibnamefont{Kvardakov}},
  \bibinfo{author}{\bibfnamefont{I.~A.} \bibnamefont{Likhachev}},
  \bibinfo{author}{\bibfnamefont{O.~V.} \bibnamefont{Vihrova}},
  \bibnamefont{et~al.}, \textbf{\bibinfo{volume}{107}}, \bibinfo{pages}{023905}
  (\bibinfo{year}{2010}), ISSN \bibinfo{issn}{00218979},
  \urlprefix\url{http://dx.doi.org/10.1063/1.3267314}.

\bibitem[{\citenamefont{Dorokhin et~al.}(2010)\citenamefont{Dorokhin, Zaitsev,
  Brichkin, Vikhrova, Danilov, Zvonkov, Kulakovskii, Prokof'eva, and
  Sholina}}]{springerlink:10.1134/S1063783410110144}
\bibinfo{author}{\bibfnamefont{M.}~\bibnamefont{Dorokhin}},
  \bibinfo{author}{\bibfnamefont{S.}~\bibnamefont{Zaitsev}},
  \bibinfo{author}{\bibfnamefont{A.}~\bibnamefont{Brichkin}},
  \bibinfo{author}{\bibfnamefont{O.}~\bibnamefont{Vikhrova}},
  \bibinfo{author}{\bibfnamefont{Y.}~\bibnamefont{Danilov}},
  \bibinfo{author}{\bibfnamefont{B.}~\bibnamefont{Zvonkov}},
  \bibinfo{author}{\bibfnamefont{V.}~\bibnamefont{Kulakovskii}},
  \bibinfo{author}{\bibfnamefont{M.}~\bibnamefont{Prokof'eva}},
  \bibnamefont{and} \bibinfo{author}{\bibfnamefont{A.}~\bibnamefont{Sholina}},
  \bibinfo{journal}{Physics of the Solid State} \textbf{\bibinfo{volume}{52}},
  \bibinfo{pages}{2291} (\bibinfo{year}{2010}), ISSN \bibinfo{issn}{1063-7834},
  \bibinfo{note}{10.1134/S1063783410110144},
  \urlprefix\url{http://dx.doi.org/10.1134/S1063783410110144}.

\bibitem[{\citenamefont{Zaitsev et~al.}(2010)\citenamefont{Zaitsev, Dorokhin,
  Brichkin, Vikhrova, Danilov, Zvonkov, and
  Kulakovskii}}]{springerlink:10.1134/S0021364009220056}
\bibinfo{author}{\bibfnamefont{S.}~\bibnamefont{Zaitsev}},
  \bibinfo{author}{\bibfnamefont{M.}~\bibnamefont{Dorokhin}},
  \bibinfo{author}{\bibfnamefont{A.}~\bibnamefont{Brichkin}},
  \bibinfo{author}{\bibfnamefont{O.}~\bibnamefont{Vikhrova}},
  \bibinfo{author}{\bibfnamefont{Y.}~\bibnamefont{Danilov}},
  \bibinfo{author}{\bibfnamefont{B.}~\bibnamefont{Zvonkov}}, \bibnamefont{and}
  \bibinfo{author}{\bibfnamefont{V.}~\bibnamefont{Kulakovskii}},
  \bibinfo{journal}{JETP Letters} \textbf{\bibinfo{volume}{90}},
  \bibinfo{pages}{658} (\bibinfo{year}{2010}), ISSN \bibinfo{issn}{0021-3640},
  \bibinfo{note}{10.1134/S0021364009220056},
  \urlprefix\url{http://dx.doi.org/10.1134/S0021364009220056}.

\bibitem[{\citenamefont{Schneider et~al.}(1987)\citenamefont{Schneider,
  Kaufmann, Wilkening, Baeumler, and K\"ohl}}]{PhysRevLett.59.240}
\bibinfo{author}{\bibfnamefont{J.}~\bibnamefont{Schneider}},
  \bibinfo{author}{\bibfnamefont{U.}~\bibnamefont{Kaufmann}},
  \bibinfo{author}{\bibfnamefont{W.}~\bibnamefont{Wilkening}},
  \bibinfo{author}{\bibfnamefont{M.}~\bibnamefont{Baeumler}}, \bibnamefont{and}
  \bibinfo{author}{\bibfnamefont{F.}~\bibnamefont{K\"ohl}},
  \bibinfo{journal}{Phys. Rev. Lett.} \textbf{\bibinfo{volume}{59}},
  \bibinfo{pages}{240} (\bibinfo{year}{1987}),
  \urlprefix\url{http://link.aps.org/doi/10.1103/PhysRevLett.59.240}.

\bibitem[{\citenamefont{Averkiev and Il'inskii}(1994)}]{Averkiev}
\bibinfo{author}{\bibfnamefont{N.~S.} \bibnamefont{Averkiev}} \bibnamefont{and}
  \bibinfo{author}{\bibfnamefont{S.~Y.} \bibnamefont{Il'inskii}},
  \bibinfo{journal}{Phys. Solid. State} \textbf{\bibinfo{volume}{36}},
  \bibinfo{pages}{278} (\bibinfo{year}{1994}).

\bibitem[{\citenamefont{Bardeen}(1961)}]{PhysRevLett.6.57}
\bibinfo{author}{\bibfnamefont{J.}~\bibnamefont{Bardeen}},
  \bibinfo{journal}{Phys. Rev. Lett.} \textbf{\bibinfo{volume}{6}},
  \bibinfo{pages}{57} (\bibinfo{year}{1961}),
  \urlprefix\url{http://link.aps.org/doi/10.1103/PhysRevLett.6.57}.

\bibitem[{\citenamefont{Schairer and Schmidt}(1974)}]{PhysRevB.10.2501}
\bibinfo{author}{\bibfnamefont{W.}~\bibnamefont{Schairer}} \bibnamefont{and}
  \bibinfo{author}{\bibfnamefont{M.}~\bibnamefont{Schmidt}},
  \bibinfo{journal}{Phys. Rev. B} \textbf{\bibinfo{volume}{10}},
  \bibinfo{pages}{2501} (\bibinfo{year}{1974}),
  \urlprefix\url{http://link.aps.org/doi/10.1103/PhysRevB.10.2501}.

\bibitem[{\citenamefont{Sapega et~al.}(2006)\citenamefont{Sapega, Ramsteiner,
  Brandt, D\"aweritz, and Ploog}}]{PhysRevB.73.235208}
\bibinfo{author}{\bibfnamefont{V.~F.} \bibnamefont{Sapega}},
  \bibinfo{author}{\bibfnamefont{M.}~\bibnamefont{Ramsteiner}},
  \bibinfo{author}{\bibfnamefont{O.}~\bibnamefont{Brandt}},
  \bibinfo{author}{\bibfnamefont{L.}~\bibnamefont{D\"aweritz}},
  \bibnamefont{and} \bibinfo{author}{\bibfnamefont{K.~H.} \bibnamefont{Ploog}},
  \bibinfo{journal}{Phys. Rev. B} \textbf{\bibinfo{volume}{73}},
  \bibinfo{pages}{235208} (\bibinfo{year}{2006}),
  \urlprefix\url{http://link.aps.org/doi/10.1103/PhysRevB.73.235208}.

\bibitem[{\citenamefont{Jungwirth et~al.}(2006)\citenamefont{Jungwirth, Sinova,
  Ma\ifmmode~\check{s}\else \v{s}\fi{}ek, Ku\ifmmode~\check{c}\else
  \v{c}\fi{}era, and MacDonald}}]{RevModPhys.78.809}
\bibinfo{author}{\bibfnamefont{T.}~\bibnamefont{Jungwirth}},
  \bibinfo{author}{\bibfnamefont{J.}~\bibnamefont{Sinova}},
  \bibinfo{author}{\bibfnamefont{J.}~\bibnamefont{Ma\ifmmode~\check{s}\else
  \v{s}\fi{}ek}},
  \bibinfo{author}{\bibfnamefont{J.}~\bibnamefont{Ku\ifmmode~\check{c}\else
  \v{c}\fi{}era}}, \bibnamefont{and} \bibinfo{author}{\bibfnamefont{A.~H.}
  \bibnamefont{MacDonald}}, \bibinfo{journal}{Rev. Mod. Phys.}
  \textbf{\bibinfo{volume}{78}}, \bibinfo{pages}{809} (\bibinfo{year}{2006}),
  \urlprefix\url{http://link.aps.org/doi/10.1103/RevModPhys.78.809}.

\bibitem[{\citenamefont{Woodbury and Blakemore}(1973)}]{PhysRevB.8.3803}
\bibinfo{author}{\bibfnamefont{D.~A.} \bibnamefont{Woodbury}} \bibnamefont{and}
  \bibinfo{author}{\bibfnamefont{J.~S.} \bibnamefont{Blakemore}},
  \bibinfo{journal}{Phys. Rev. B} \textbf{\bibinfo{volume}{8}},
  \bibinfo{pages}{3803} (\bibinfo{year}{1973}),
  \urlprefix\url{http://link.aps.org/doi/10.1103/PhysRevB.8.3803}.

\end{thebibliography}

%\pagebreak

\end{document}